\documentstyle[epsfig]{aipproc}

\def\Journal#1#2#3#4{{#1} {\bf #2}, #3 (#4)}

\def\PLB{{\em Phys. Lett.}  B}

\def\be{\begin{equation}}
\def\ee{\end{equation}}
\def\bea{\begin{eqnarray}}
\def\eea{\end{eqnarray}}

\begin{document}
\title{Fermion and Higgs Masses and the AGUT Model}

\author{Colin D. Froggatt$^*$ and Holger B. Nielsen$^{\dagger}$}
\address{$^*$Niels Bohr Institute, Copenhagen {\O},
Denmark\\
$^{\dagger}$Department of Physics and Astronomy,
Glasgow University, Glasgow G12 8QQ, Scotland}

\maketitle

\begin{abstract}
We present two rather differently based predictions for the
quark and lepton spectrum: One provides a
rather successful fit
to the mass suppressions---the well known
fermion mass hierarchy---interpreted
as due to most mass terms needing to
violate approximately conserved quantum
numbers corresponding to the AGUT group $ SMG^3\times U(1)_f$. This is
actually, under certain conditions, the maximal group
transforming the known 45 Weyl
components of the quark and leptons into each other.
{}From the fit to the fermion spectrum,
we get a picture of the series of Higgs fields
causing the breakdown (presumably at the Planck scale) of this
AGUT to the Standard Model and, thus, providing the small masses of
all quarks and leptons except for the top quark.
We separately predict the top quark mass to be
$173 \pm 5$ GeV and the Higgs mass to be
$135 \pm 9$ GeV, from the assumption
that there be two degenerate minima in the
effective potential for the
Weinberg Salam Higgs field with the second one
at the Planck field strength.

\end{abstract}

\section{Introduction}\label{sec:intro}

All of the charged fermion masses, apart from the top quark,
are suppressed relative to the electroweak scale. Indeed the
most striking feature of their spectrum is the hierarchy of masses
and mixing matrix elements: the masses range over five orders
of magnitude, from 1/2 MeV for the electron to 175 GeV for the
top quark. The most promising way of explaining this hierarchy is
in terms of mass suppression factors, due to the partial conservation
of some chiral flavour quantum numbers beyond the Standard Model (SM).
In section \ref{sec:mass} we consider a
model \cite{smg3m} based on the Anti-Grand
Unified Theory (AGUT) gauge group $SMG^3\times U(1)_f$.
This means that, near the Planck scale
$M_{Planck} \simeq 10^{19}$ GeV, each quark-lepton
generation has its own set of gauge fields quite analogous
to those of the Standard Model
Group $SMG = SU(3) \times SU(2) \times U(1)$. In addition
there is an extra abelian gauge group called $U(1)_f$.
We shall characterize the $SMG^3\times U(1)_f$ group
as the maximal AGUT group, satisfying some relatively
simple assumptions, in section \ref{sec:max}.
This gauge group is supposed to break down to the
Standard Model Group $SMG$, as the diagonal subgroup of
$SMG^3$, about one order of magnitude below $M_{Planck}$.
The vacuum expectation values (VEVs), measured in units of
$M_{Planck}$, of the Higgs fields
responsible for the breakdown provide the required
fermion mass suppression factors.

In section \ref{sec:top} we present a precise determination
of the top quark and Higgs boson masses within the pure
SM, based on the principle of degenerate vacua
and a strongly first order phase transition \cite{smtop}.
The values of the
Yukawa coupling constant $g_t$ and the Higgs self-coupling
$\lambda$ then take on fine-tuned values, rather analogous
to a mixture of ice and water taking on a fine-tuned temperature
equal to zero degrees Celsius. In winter one often finds
a mixture---slush---of ice and water. This is due to the
large latent heat of water giving a strongly first order
phase transition between ice and water. We assume that
the SM is valid up close to the Planck scale;
the strongly first order phase transition condition
then implies that the SM effective Higgs potential
should have two degenerate minima, with the corresponding
difference in Higgs field VEVs being of order $M_{Planck}$.

\section{The Maximal Group} \label{sec:max}

The $SMG^3\times U(1)_f$ group, with its 37 generators, at
first seems a rather arbitrary choice for a ``unified group".
However it can be characterized uniquely as the gauge group $G$
beyond the SM (i.e. having SMG as a subgroup) satisfying the
following 4 postulates:
\begin{enumerate}
\item $G \subseteq U(45)$. Here $U(45)$ is the group of all
unitary transformations of the 45 species of Weyl fields (3
generations with 15 in each) in the SM.
\item No anomalies. There should be neither gauge anomalies nor
mixed anomalies. We assume that only straightforward anomaly
cancellation takes place and, as in the SM itself, do not allow
for a Green-Schwarz type anomaly cancellation \cite{green}.
\item The various irreducible representations of Weyl fields
for the SMG remain irreducible under $G$. This
postulate is motivated by the observation that combining SM
irreducible representations into larger unified representations
introduces symmetry relations between Yukawa coupling constants,
whereas the particle spectrum exhibits a hierarchy between
essentially all the fermion masses rather than exact degeneracies.
\item $G$ is the maximal group satisfying the other 3 postulates.
\end{enumerate}

A rather complicated calculation shows that, modulo
permutations of the various SM fermion irreducible representations,
we are led to the result $G = SMG^3 \times U(1)_f$ with the usual
SMG embedded as the diagonal subgroup of $SMG^3$. Apart from
the various permutations of the particle names, the $U(1)_f$
group is unique. The $U(1)_f$ charges $Q_f$
can then be chosen so that the only non-zero values are carried
by the right-handed fermions of the second and third
proto-generations:
\begin{equation}
Q_f(\tau_R) = Q_f(b_R) = Q_f(c_R) = 1
\quad Q_f(\mu_R) = Q_f(d_R) = Q_f(t_R) = -1
\end{equation}

\section{The Fermion Mass Spectrum}\label{sec:mass}

The Yukawa couplings of the SM fermions to the
Weinberg-Salam Higgs field $\phi_{WS}$ are mainly forbidden by the
gauge quantum numbers of the AGUT group. In the AGUT theory,
these transitions between the left- and right-handed Weyl
fields also involve the Higgs fields responsible for the
breakdown of the $SMG^3 \times U(1)_f$ group. We assume that all the
fundamental couplings of the AGUT theory are of order unity and
that there exists a rich spectrum of vector-like fermion states
at the Planck scale, which can mediate all the required
transitions.
Then the gauge quantum numbers of the quark-lepton and
Higgs fields determine the combinations of Higgs fields needed
to provide non-zero values for the various elements in the
effective SM Yukawa coupling matrices $Y_U$, $Y_D$ and $Y_E$
for the up quarks, down quarks and charged leptons respectively.
We have the freedom to choose the quantum numbers of the
Higgs fields and, guided by phenomenology, we select
four Higgs fields S, W, T and $\xi$, in addition to $\phi_{WS}$.
In fact we specify their U(1) charges and use a natural
generalisation of the  SM charge quantisation rule to determine
their non-abelian representations.
The Higgs field S is supposed
to have a VEV of order unity in Planck units and,
therefore, does not
contribute to the fermion mass suppression. In this way we
find the following order of magnitude SM Yukawa coupling
matrices:
\begin{eqnarray}
Y_U & \simeq &
\left ( \begin{array}{ccc}
	S^{\dagger}W^{\dagger}T^2(\xi^{\dagger})^2 & W^{\dagger}T^2\xi &
		(W^{\dagger})^2T\xi \\
	S^{\dagger}W^{\dagger}T^2(\xi^{\dagger})^3 & W^{\dagger}T^2 &
		(W^{\dagger})^2T \\
	S^{\dagger}(\xi^{\dagger})^3 & 1 & W^{\dagger}T^{\dagger}
			\end{array} \right ) \label{H_U}
\\
Y_D & \simeq &
\left ( \begin{array}{ccc}
	SW(T^{\dagger})^2\xi^2 & W(T^{\dagger})^2\xi & T^3\xi \\
	SW(T^{\dagger})^2\xi & W(T^{\dagger})^2 & T^3 \\
	SW^2(T^{\dagger})^4\xi & W^2(T^{\dagger})^4 & WT
			\end{array} \right ) \label{H_D}
\\
Y_E & \simeq &
\left ( \begin{array}{ccc}
	SW(T^{\dagger})^2\xi^2 & W(T^{\dagger})^2(\xi^{\dagger})^3 &
		(S^{\dagger})^2WT^4\xi^{\dagger} \\
	SW(T^{\dagger})^2\xi^5 & W(T^{\dagger})^2 &
	(S^{\dagger})^2WT^4\xi^2 \\
	S^3W(T^{\dagger})^5\xi^3 & (W^{\dagger})^2T^4 & WT
			\end{array} \right ) \label{H_E}
\end{eqnarray}
where the Higgs fields are replaced by their VEVs in Planck units.
\begin{table}[h]
\caption{Best fit to experimental data.
All masses are running masses at 1 GeV
except the top quark mass which is the pole mass.}
\label{bestfit}
\begin{tabular}{ccccccc}
 & $M_t$ & $m_b$ & $m_{\tau}$  & $m_c$ & $m_s$ & $m_{\mu}$ \\
\tableline
Fit & 192 GeV & 8.3 GeV &
1.27 GeV  & 1.02 GeV & 400 MeV &
88 MeV \\
Data  & 180 GeV & 6.3 GeV &
1.78 GeV & 1.4 GeV & 200 MeV &
105 MeV \\
\end{tabular}
\end{table}
\begin{table}
\hspace{2 cm}
\begin{tabular}{ccccccc}
& $m_u$ & $m_d$ & $m_e$ & $V_{us}$ & $V_{cb}$ & $V_{ub}$ \\
\tableline
Fit  & 3.6 MeV & 7.0 MeV &
0.87 MeV & 0.18 & 0.018 & 0.0039 \\
Data  & 4 MeV & 9 MeV &
0.5 MeV & 0.22 & 0.041 & 0.0035 \\
\end{tabular}
\end{table}

The diagonal elements in all 3 Yukawa matrices
have the same form, up to
complex conjugation, giving the order of magnitude SU(5)-like
results $m_b \approx m_{\tau}$ and $m_s \approx m_{\mu}$, but the
off-diagonal elements dominate $Y_U$ making $m_t$ and $m_c$
respectively larger. The diagonal and off-diagonal contributions to
the lowest eigenvalue of $Y_D$ are approximately equal, giving
$m_d \gtrsim m_u \approx m_e$. A three parameter
order of magnitude fit \cite{smg3m},
including random complex factors of order unity,
with $S = 1$ fixed,
$\langle W \rangle = 0.179$, $\langle T \rangle =0.071$ and
$\langle \xi \rangle = 0.099$ successfully reproduces the
9 masses and 3 mixing angles---see Table \ref{bestfit}.

\section{Top Quark and Higgs Masses}\label{sec:top}

We now apply our principle of degenerate vacua
and the strongly first order phase transition requirement to
the pure SM, with a desert up to the Planck
scale \cite{smtop}. It is well-known that, with loop corrections, the SM
effective Higgs potential can have two minima. So we are led
to our two crucial assumptions:
\newline
a) The two minima in the Standard Model effective Higgs potential
are degenerate:
$V_{\rm{eff}}(\phi_{\rm{min}\; 1}) = V_{\rm{eff}}(\phi_{\rm{min} \; 2})$.
\newline
b) The second minimum has a
Higgs field squared of the order of unity
in Planck units:
$<|\phi_{\rm{min} \; 2}|^2> = {\cal O}(M_{Planck}^2)
\sim (10^{-19}$ GeV)$^2$.

\begin{figure}[h]
\leavevmode
\centerline{
\epsfig{file=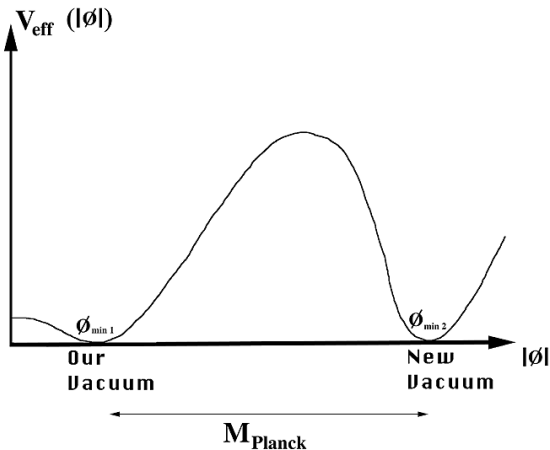,width=8.5cm,%
bbllx=0pt,bblly=0pt,bburx=548pt,bbury=444pt,%
clip=}
}
\caption{
This symbolic graph of the effective potential
$V_{\rm{eff}}(\phi)$ for the Standard Model Higgs field
illustrates the two assumptions which lead to our prediction of
the top quark and Higgs boson masses:
1) Two equally deep minima, 2) achieved for $|\phi|$
values differing, order of magnitudewise, by
unity in Planck units.
}
\label{fig:veff}
\end{figure}

We use the renormalisation group improved tree level effective potential,
identifying the renormalisation point with the field strength $\phi$:
\begin{equation}
V_{\rm{eff}}(\phi) \; = \; \frac{1}{2}m_{h}^2(\mu = |\phi |)\,
|\phi |^2 \; + \; \frac{1}{8}\lambda (\mu = |\phi | )\, |\phi |^4
\end{equation}
The condition $V_{\rm{eff}}(\phi_{\rm{min}\; 1})
= V_{\rm{eff}}(\phi_{\rm{min} \; 2})$,
where one of the minima corresponds to our vacuum with
$\phi_{\rm{min}\; 1} = 246$ GeV,
defines (part of) the well-known
vacuum stability curve in the $M_t-M_H$ plane,
for $\phi_{\rm{min} \; 2} < M_{Planck}$.
For values of $\phi$ of order $M_{Planck}$,
the $|\phi|^4$ term
dominates the $|\phi|^2$ term
and the vacuum degeneracy condition requires
$\lambda(\phi_{\rm{min} \; 2})$ to vanish. Also the derivative
of $V_{\rm{eff}}(\phi)\approx
\frac{1}{8}\lambda(\phi) \phi^4 $ should be zero:
\begin{equation}
\frac{dV_{\rm{eff}}(\phi)}{d\phi}|_{\phi_{\rm{min} \; 2}}
= \frac{1}{2}\lambda(\phi)\phi^3
+\frac{1}{8}\frac{d\lambda(\phi)}{d\phi}\phi^4
=\frac{1}{8}\beta_{\lambda} \phi^3 = 0
\end{equation}
and thus the beta function $\beta_{\lambda}(\phi_{\rm{min} \; 2})$
vanishes as well. So we impose the conditions
$\beta_{\lambda}=\lambda=0$
near the Planck scale and, using the renormalisation
group equations, determine a single point on the
SM vacuum stability curve. In this way our two assumptions,
illustrated in Figure \ref{fig:veff}, lead to our predictions for the
top quark and Higgs boson pole masses:
\begin{equation}
M_{t} = 173 \pm 4\ \mbox{GeV} \quad M_{H} = 135 \pm 9\ \mbox{GeV}.
\end{equation}

\section{Conclusion}\label{sec:con}

We have found a rather good fit, that in principle should only work
to order of magnitude accuracy,
for the nine charged quark and lepton masses
and the three mixing angles; it
even fits the CP-violation reasonably well.
Three suppression factors were used, indentified in the
model with expectation values of three Higgs
fields, T, W, and $\xi$.
There was also a Higgs field $S$ causing no
suppression.
Even the overall mass scale could be thought of
as being correctly predicted, in as far as it is part of our model that
unforbidden Yukawa couplings are of order of magnitude unity
and therefore the top quark mass corresponds to the
electroweak scale.
Since our fit is a priori based on the assumption of the
AGUT gauge group,
$SMG^3\times U(1)_f$, one may at first think that we could take the
goodness of our fit as evidence for
this gauge group really being realized
at some very high energy scale---say the Planck scale,
to which we actually extrapolated in our detailed fit.
However,
first the field $S$ caused a breakdown of this group to the
subgroup $SMG^2\times U(1)$ = $ SMG_{12}\times SMG_3 \times U(1)$.
Thus our model for the fermion masses really
only used the quantum numbers of this subgroup and checked for its
presence in the gauge group.
Secondly we did not have to use the non-abelian parts of the
group, but obtained our results using just the U(1) factors.
So concluding back to the relevance of the proposed
AGUT group is somewhat doubtful.

However, we also pointed out that our $SMG^3\times U(1)_f$ gauge group
could be specified rather simply by means of four suggestive postulates.
Thus when this relatively easy
to characterize group, $SMG^3\times U(1)_f,$ turns out to provide a
consistent mass matrix fit,
it is rather suggestive after all
that it is indeed the correct gauge group.

An at first rather unrelated calculation gave us the
top quark mass with good accuracy $M_t = 173 \pm 5$ GeV, and not only its
order of magnitude as in the just mentioned AGUT fit.
We required the Weinberg-Salam Higgs potential
of the pure Standard Model to have
two degenerate minima, one being at the Planck scale, as in
Figure \ref{fig:veff}.
{}From the same requirement, the Higgs mass is predicted
to have that value which barely allows the
stability of the SM vacuum: $M_H = 135 \pm 9$ GeV.

The two calculations are actually connected,
in as far as our fine structure constant predictions, described in
Don Bennett's talk \cite{don},
are based on the assumptions underlying both calculations:
the AGUT gauge group (ignoring though the $U(1)_f$) and the principle
that there shall be degenerate vacua.

It should be stressed that the more precise top
quark mass prediction was
based on the assumption that there should
be a desert almost up to the Planck
scale.
So it would be a falsification of the simplest---and presumably the
only sensible---version of our model, if new physics
particles such as, for example, SUSY particles are
found, which would give so strong
contributions to the renormalisation group running
for the top and Higgs couplings that it would disturb our top
quark mass prediction.

However, our order of magnitude fit to the quark and lepton spectrum
is not very sensitive to the scale
at which our proposed AGUT gauge group
should be found. It is true that we extrapolate
to the Planck scale and thereby
use the simple suppression factors to fit the Planck scale
Yukawa couplings. This extrapolation essentially just provides
each lepton mass with an extra factor 3 to 4 relative to the
quark masses, because the lepton Yukawa couplings are
running less than the quark Yukawas. But a factor 3 to 4 is not
sensitively tested when we only work with orders of magnitude.

\end{document}